\newcommand{\fig}[1]{\mbox{Fig.\hspace{0.2em}\ref{#1}}}
\newcommand{\eqn}[1]{\mbox{Eq.\hspace{0.2em}\ref{#1}}}
\newcommand{\sect}[1]{\mbox{\S\ref{#1}}}

\newcommand{\aeff}{\Omega_\mathrm{e}}
\newcommand{\vrel}{v_\mathrm{rel}}
\newcommand{\np}{N_\mathrm{P}}
\newcommand{\nt}{N_\mathrm{T}}
\newcommand{\nexp}{N_\mathrm{exp}}

\documentclass{emulateapj}
\usepackage{graphicx}
\usepackage{natbib}
\usepackage{amssymb}

\begin{document}

\title{First Results From The Taiwanese-American Occultation Survey (TAOS)}
\author{
Z.-W.~Zhang\altaffilmark{1},
F.~B.~Bianco\altaffilmark{2,3},
M.~J.~Lehner\altaffilmark{4,3},
N.~K.~Coehlo\altaffilmark{5},
J.-H.~Wang\altaffilmark{1,4},
S.~Mondal\altaffilmark{1},
C.~Alcock\altaffilmark{3},
T.~Axelrod\altaffilmark{6},
Y.-I.~Byun\altaffilmark{7},
W.~P.~Chen\altaffilmark{1},
K.~H.~Cook\altaffilmark{8},
R.~Dave\altaffilmark{9},
I.~de~Pater\altaffilmark{10},
R.~Porrata\altaffilmark{10},
D.-W.~Kim\altaffilmark{7},
S.-K.~King\altaffilmark{4},
T.~Lee\altaffilmark{4},
H.-C.~Lin\altaffilmark{1},
J.~J.~Lissauer\altaffilmark{11},
S.~L.~Marshall\altaffilmark{12,8},
P.~Protopapas\altaffilmark{3},
J.~A.~Rice\altaffilmark{5},
M.~E.~Schwamb\altaffilmark{13},
S.-Y.~Wang\altaffilmark{4} and
C.-Y.~Wen\altaffilmark{4}
}
\altaffiltext{1}{Institute of Astronomy, National Central University, No. 300,
 Jhongda Rd, Jhongli City, Taoyuan County 320, Taiwan}
\email{s1249001@cc.ncu.edu.tw}
\altaffiltext{2}{Department of Physics and Astronomy, University of
 Pennsylvania, 209 South 33rd Street, Philadelphia, PA 19104}
\altaffiltext{3}{Harvard-Smithsonian Center for Astrophysics, 60 Garden Street,
 Cambridge, MA 02138}
\altaffiltext{4}{Institute of Astronomy and Astrophysics, Academia Sinica.
 P.O. Box 23-141, Taipei 106, Taiwan}
\altaffiltext{5}{Department of Statistics, University of California Berkeley,
 367 Evans Hall, Berkeley, CA 94720}
\altaffiltext{6}{Steward Observatory, 933 North Cherry Avenue, Room N204
 Tucson AZ 85721}
\altaffiltext{7}{Department of Astronomy, Yonsei University, 134 Shinchon,
 Seoul 120-749, Korea}
\altaffiltext{8}{Institute of Geophysics and Planetary Physics, Lawrence
 Livermore National Laboratory, Livermore, CA 94550}
\altaffiltext{9}{Initiative in Innovative Computing, Harvard University,
 60 Oxford St, Cambridge, MA 02138}
\altaffiltext{10}{Department of Astronomy, University of California Berkeley,
 601 Campbell Hall, Berkeley CA 94720}
\altaffiltext{11}{Space Science and Astrobiology Division 245-3,
 NASA Ames Research Center, Moffett Field, CA, 94035}
\altaffiltext{12}{Kavli Institute for Particle Astrophysics and Cosmology,
 2575 Sand Hill Road, MS 29, Menlo Park, CA 94025}
\altaffiltext{13}{Division of Geological and Planetary Sciences,
 California Institute of Technology, 1201 E. California Blvd., Pasadena, CA
 91125}

\begin{abstract}
Results from the first two years of data from the Taiwanese-American
Occultation Survey (TAOS) are presented. Stars have been monitored
photometrically at 4~Hz or 5~Hz to search for occultations by small
($\sim$3~km) Kuiper Belt Objects (KBOs).  No statistically significant
events were found, allowing us to present an upper bound to the size
distribution of KBOs with diameters 0.5~km~$<D<$~28~km.
\end{abstract}

\keywords{Kuiper Belt, occultations, solar system: formation}

\section{Introduction}
\setcounter{footnote}{0}

The study of the Kuiper Belt has exploded since the discovery of 1992
QB1 by \citet{1993Natur.362..730J}.  The brightness distribution of
objects with $R$ magnitude brighter than $\sim$26 is relatively
well-established by many surveys, most recently by \citet[and
  references therein]{2008Icar..195..827F} .  The brightness
distribution is adequately described by a simple cumulative luminosity
function \mbox{$\Sigma(<R) = 10^{\alpha(R-R_0)}$~deg$^{-2}$}, where
$R_0 \sim 23$ and $\alpha \sim 0.6$, for objects with magnitude
$R<26$.  There is clear evidence for a break to a shallower slope for
fainter objects: the deepest survey, conducted using the Advanced
Camera for Surveys on the \emph{Hubble Space Telescope}
\citep{2004AJ....128.1364B} extended to $R = 28.5$, and found a factor
of $\sim$25 fewer objects than would be expected if the same
distribution extended into this range.

The size distribution of Kuiper Belt Objects (KBOs) is believed to
reflect a history of \emph{agglomeration} during the planetary formation
epoch, when relative velocities between particles were low and
collisions typically resulted in particles sticking together, followed
by \emph{destructive collisions} when the relative velocities were
increased by dynamical processes after the giant planets formed
\citep{1996AJ....112.1203S, 1997Icar..125...50D, 1997AJ....114..841S,
  1999AJ....118.1101K, 1999ApJ...526..465K, 2004AJ....128.1916K,
  2005Icar..173..342P}. The slope of the distribution function for
larger objects reflects the early phase of agglomeration, while the
shallower distribution for smaller objects reflects a subsequent phase
of destructive collisions. The location of the break moves to larger
sizes with time, while the distribution for smaller objects is
expected to evolve towards a steady state collisional cascade
\citep{2004AJ....128.1916K, 2005Icar..173..342P}. Models for the
spectrum of small bodies differ between \citet{2005Icar..173..342P},
who derived a double power-law distribution, and
\citet{2004AJ....128.1916K}, whose simulations show more structure,
depending on material properties.


Thus, the size spectrum encodes information about the history of
planet formation and dynamics. However, the size spectrum for small
KBOs is not constrained by the imaging surveys because the objects of
interest are too faint for direct detection using presently available
instruments. These small objects may, however, be detected indirectly
when they pass between an observer and a distant star
\citep{1976Natur.259..290B, 1992QJRAS..33...45D, 1992ASPC...34..171A,
  1997MNRAS.289..783B, 2000Icar..147..530R, 2003ApJ...587L.125C,
  2007AJ....134.1596N}. The challenge confronting any survey
exploiting this technique is the combination of very low anticipated
event rate and short duration of the events (typically $<$ 1~second).

\begin{figure*}[bt]
\plottwo{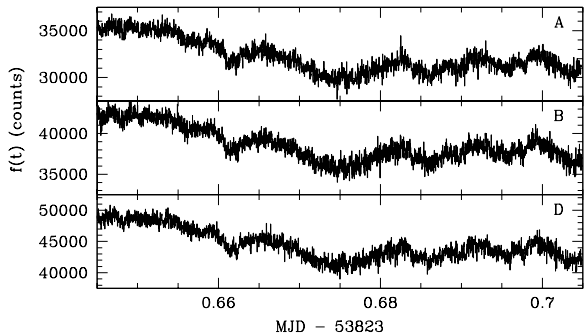}{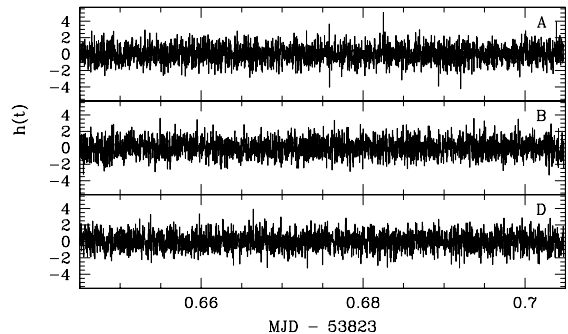}
\caption[]{Demonstration of detrending by the filtering process. The
  left panel shows a raw lightcurve set $f(t)$, and the right panel
  shows the lightcurve set $h(t)$ after filtering. The total number
  points in the lightcurve set is 26,622, corresponding to
  89~minutes. Here only one of every 10~points is plotted for
  readability.}
\label{fig:lcdemo}
\end{figure*}

Other groups are attempting similar occultation
surveys. \citet{2006AJ....132..819R} reported three events in
10~star-hours of photometric data sampled at 45~Hz, which they modeled
as objects at 15~AU, 140~AU, and 210~AU, respectively, placing the
inferred objects outside the Kuiper Belt.  \citet{2008AJ....135.1039B}
reported results of 5~star-hours of data sampled at 40~Hz, during
which no events were detected.  \citet{2006Natur.442..660C} reported a
surprisingly high rate of possible occultation events in \emph{RXTE}
x-ray observations of Sco X-1, but many of these events have since
been attributed to instrumental effects \citep{2008ApJ...677.1241J,
  2007MNRAS.378.1287C}.

We report here the first results of the Taiwanese American Occultation
Survey (TAOS). TAOS differs from the previously reported projects
primarily in the extent of the photometric time series, a total of
$1.53 \times 10^5$~star-hours, and in that data are collected
simultaneously with three telescopes. Some compromises have been made
in regard to signal-to-noise (SNR), which is typically lower than in
previously reported surveys, and in cadence, which is 4~Hz or 5~Hz, in
contrast to the higher rates mentioned above.  The substantial
increase in exposure more than compensates for the lower cadence and
SNR, and we are able to probe significant ranges of the model space
for small KBOs.  We have also developed a statistical analysis
technique which allows efficient use of the multi-telescope data to
detect brief occultation events that would be statistically
insignificant if observed with only one telescope.

\section{Data and Analysis}
\label{sec:analysis}

TAOS has been collecting scientific data since 2005. Observations are
normally carried out simultaneously with three 50 cm telescopes (A, B,
and D, separated by distances of 6 meters and 60 meters; this system
is described by \citet{2008arXiv0802.0303L}).  Over 15~TB of raw
images have been taken.  We report here on the first two years of data
taken simultaneously with all three telescopes. The data set comprises
156~\emph{data runs}, where a data run is defined as a series of
three-telescope observations of a given field for durations of
$\sim$90~minutes. Thirty data runs with a 4~Hz sampling rate were
taken before 2005 December 15, and 126~data runs were collected
subsequently with a sampling rate of 5~Hz.  Only fields with ecliptic
latitudes $|b| < 10^\circ$ were analyzed. Over 93\% of the data were
collected in fields with $|b| < 3^\circ$, so the results of our
analysis are relevant to the sum of the cold and excited KBO
populations \citep{2004AJ....128.1364B}. No data run was included
unless each star was sampled more than 10,000~times in each telescope.
The angle from opposition in these data runs is distributed from
$0^\circ$ to $90^\circ$. The number of stars (with $R< 13.5$, which
typically gives a SNR $\ge 5$) monitored in the data runs ranges
between 200 and 2000\footnote{Information on the TAOS fields is
  available at \url{http://taos.asiaa.sinica.edu.tw/taosfield/}.}.

The images were analyzed using an aperture photometry package
\citep{kiwi} devised exclusively for TAOS images.  Lightcurves were
produced for each star by assembling the photometric information into
a time series.  A star in each data run has a lightcurve from each of
the three telescopes.  The data presented in this paper comprises
110,895 \emph{lightcurve sets} (where a lightcurve set is defined as a
set of three lightcurves, one for each telescope, for the same star in
a single data run), containing $7.1 \times 10^9$ individual
photometric measurements.

The photometric data are not calibrated to a standard system.  Changes
in atmospheric transparency during a data run produce flux
variations (\fig{fig:lcdemo}) that could undermine our occultation search
algorithm.  Such low-frequency trends in a lightcurve can be removed
by a numerical high-pass filter that preserves information of a brief
occultation event, typically with a duration less than 1--2 data
points with the TAOS sampling rate.  Our filter takes a time series of
$ f_{i}$ measured at time $t_i$ to produce an intermediate series
$g_{i} = f_{i} - \bar{f_{i}}$ where $\bar{f_{i}}$ is the running
average of 33 data points centered on $t_{i}$.  The series $g_{i}$ is
then scaled by the local fluctuation, $h_{i} = g_{i} / \sigma(g_{i})$
with $\sigma(g_{i})$ being the standard deviation of $g_{i}$ of 151
data points centered at $t_{i}$.  Both the mean and standard deviation
are calculated using \emph{three-sigma clipping}.  This filtering
proves effective to remove slow-varying trends in the lightcurve,
while preserving high-frequency fluctuations that we aim to detect, as
illustrated in \fig{fig:lcdemo}.

\begin{figure}[t]
\plotone{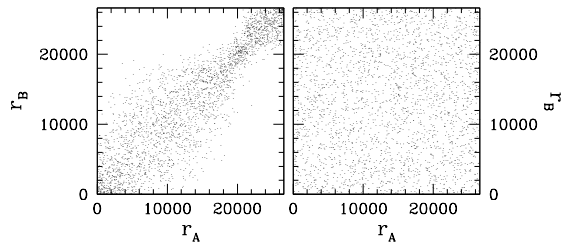}
\caption[]{Rank-rank plots of the lightcurve set shown in
  \fig{fig:lcdemo}. Each point shows the ranks at a single time point
  for telescopes A and B. The ranks from the raw lightcurves
  (\fig{fig:lcdemo}, left) are shown in the left panel, and the ranks
  from the filtered lightcurves (\fig{fig:lcdemo}, right) are shown in
  the right panel.}
\label{fig:rkdemo}
\end{figure}

We now confront the two central challenges in the search for extremely
rare occultation events in these data: \emph{(1) to search for events
  simultaneously in three parallel data streams,} and \emph{(2) to
  determine the statistical significance of any rare events that are
  found.}  The second is not straightforward because the statistical
distribution of our photometric measurements is not known in advance;
approximations based on Gaussian statistics are unreliable far from
the mean.  This motivates a non-parametric approach.

We thus found it useful to represent each data point by its
\emph{rank} in the filtered, rescaled lightcurve data $h_{i}$. That
is, the rank of a data point ranges from $r=1$ (lowest $h$) to $r=\np$
(highest $h$), for a data run comprising $\np$ photometric
measurements taken with telescope A, B, or D.  The \emph{rank triples}
$(r_{i}^A, r_{i}^B, r_{i}^D)$ form the basis of further analysis of
the multi-telescope data.  The statistical distribution of these ranks
is known exactly, since each rank must occur exactly once in each time
series. Thus, the probability that a given rank will occur at time
$t_i$ is $P(r_i)=1/\np$.  When the photometric data are uncorrelated,
the probability that a particular rank triple will occur is simply
$P(r_{i}^A, r_{i}^B, r_{i}^D)=1/\np^3$. This allows a straightforward
test for correlation between the photometric data taken in the three
telescopes: the rank triples should be distributed uniformly in a cube
with sides of length $\np$.  A non-uniform pattern in the cube, on the
other hand, indicates correlation. \fig{fig:rkdemo} shows an example
of the rank series of one telescope against another; this indicates
that the raw photometric data $f_{i}$ are strongly correlated, but the
filtered data $h_{i}$ are not.

\begin{figure}[t]
\plotone{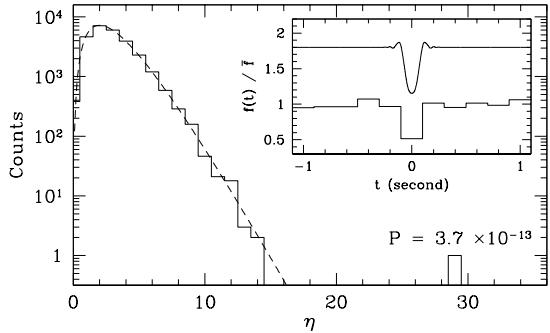}
\caption[]{Histogram of $\eta$ for a lightcurve set implanted with a
  \emph{simulated} occultation by a 3~km KBO. Each lightcurve
  comprises 26,637 points and the ranks from the three telescopes are
  found to be 1, 3, and 1. The event is clearly visible at
  $\eta=29.47$, and the probability of a rank product of 3 or lower
  from random chance is $P = 3.7\times 10^{-13}$.  The overlapped dash
  line is the theoretical distribution of $\eta$.  The theoretical and
  implanted lightcurves are shown in the insert, offset vertically for
  clarity.}
\label{fig:rhist}
\end{figure}

Given that the rank triples are uncorrelated, the ranks can be used to
search for possible occultation events, as follows: \emph{A true
  occultation event will exhibit anomalous, correlated low ranks in
  all three telescopes.}  The rank triples thus allow an elegant test
for the statistical significance of a candidate event.  Consider the
quantity
\begin{displaymath}
\eta_i = -\ln(r_{i}^{A} r_{i}^{B} r_{i}^{D} / \np^3).
\end{displaymath}
Since the ranks are uncorrelated (unless we have an occultation
event), we can calculate the exact probability density
function\footnote{For small $\eta$, this distribution can be
  approximated by a $\Gamma$~distribution of the form $P(\eta) =
  \eta^{\nt -1}\,e^{-\eta}/(\nt - 1)!$, where $\nt = 3$ is the number
  of telescopes.} for $\eta$.  This in turn allows us to compute the
probability that a given triple of low ranks randomly occurred in an
uncorrelated lightcurve set; this is our measure of statistical
significance.  An illustration of the power of this approach is shown
in \fig{fig:rhist}, where a simulated occultation event is readily
recovered.\footnote{Details of our statistical methodology will be
  described in \citet{stat}.}

We thus screened all of our series of rank triples for events with low
ranks in all three telescopes. In the analysis reported here, we
considered only events for which $P(\eta \geq c) \leq 10^{-10}$, which
leads to an expected value of $0.24$ false positive events in the
entire data set of $2.4\times 10^9$~rank triples. \emph{No
  statistically significant events emerged from this
  analysis.}\footnote{A candidate event was reported in
  \citet{2007IAUS..236...65C}.  This event had a significance of
  $3.7\times 10^{-10}$, which did not pass our cut on $\eta$. We
  expect to have $\sim$1 false positive event at that significance
  level or higher.}

\section{Efficiency Test and Event Rate}

While the example event shown in \fig{fig:rhist} is readily
recovered, objects with smaller diameters or which do not directly
cross the line of sight might not be so easily detected. An efficiency
calculation is thus necessary for understanding the detection sensitivity
of our data and analysis pipeline to different event parameters,
notably the KBO size distribution. Our efficiency test started with
implanting synthetic events into observed lightcurves, with the
original noise.  Each original lightcurve $f_{i}$ was modified by
the implanted occultation, $ k_{i} = f_{i} - (1 -d_{i})\bar{f_{i}},$
where $d_{i}$ is the simulated event lightcurve (with baseline $d_i
\rightarrow 1$ far from the event, \citep{2007AJ....134.1596N}), and
$\bar{f_{i}}$ is the average of the original series over a 33~point
rolling window.  Since we preserved the original noise in the modified
lightcurves, the noise where the implanted occultation event takes
place---for which the flux diminishes---would be slightly
overestimated, hence our efficiency estimate is conservative.

\begin{figure}[bt]
\plotone{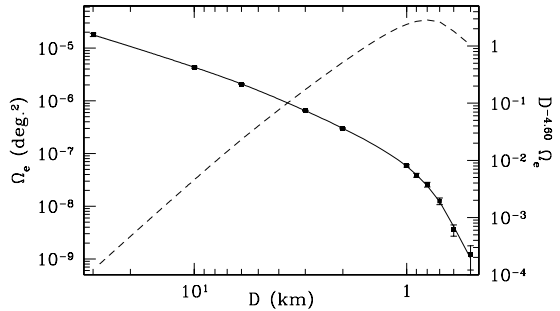}
\caption[]{Effective solid angle of the TAOS survey as a function of
  KBO diameter $D$ (solid line), and the product of the upper limit of
  the differential surface density by the effective solid angle
  (dashed line, in arbitrary units).}
\label{fig:aeff}
\end{figure}

We assumed spherical KBOs at a fixed geocentric distance of
$\Delta=43$~AU. (Given our sampling rate, varying the KBO distance
within the Kuiper Belt has little effect on our simulated lightcurves.)
The event epoch $t_0$ was chosen randomly and uniformly within the
duration $E$ of the lightcurve set. The angular size $\theta_*$ of
each star, necessary for the simulated lightcurve calculation, was
estimated using stellar color and apparent magnitude taken from the
USNO-B \citep{2003AJ....125..984M} and 2MASS
\citep{2006AJ....131.1163S} catalogs. The impact parameter of each
event was chosen, again randomly and uniformly, between 0 and $H/2$,
where $H$ is the \emph{event cross section}
\citep{2007AJ....134.1596N},
\begin{displaymath}
H = 2\left[(\sqrt{3}F)^{3 \over 2} + (D / 2)^{3 \over 2}
  \right]^{{2 \over 3}} + \theta_*\Delta.
\end{displaymath}
Here $D$ is the diameter of the occulting object, and $F$ is the
\emph{Fresnel scale}, $F = \sqrt{\lambda \Delta/2}$, where $\lambda =
600$~nm is the median wavelength in the TAOS filter. The relative
velocity between the Earth and KBO $\vrel$, necessary for the
conversion of the occultation diffraction pattern to a temporally
sampled lightcurve, is calculated based on the angle from opposition
during each data run.

\begin{figure}[bt]
\plotone{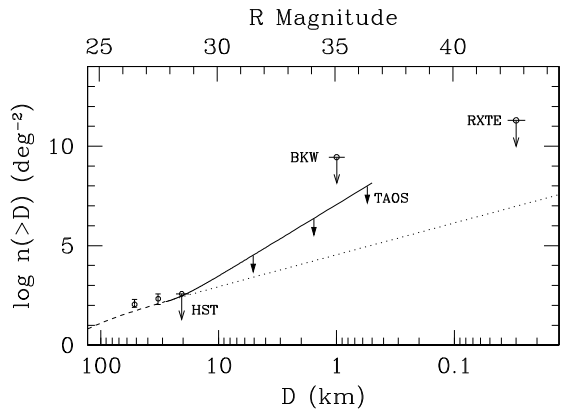}
\caption[]{TAOS upper limit to the size spectrum (solid line) assuming
  a power-law distribution; double power-law fit from
  \citet{2004AJ....128.1364B} (dashed line, extrapolation at $D<28$~km
  shown in dotted line).  Results from \citet{2004AJ....128.1364B} are
  shown as data points.  Upper limits from \citet{2008AJ....135.1039B}
  (BKW) and \citet{2008ApJ...677.1241J} (RXTE) also shown. An albedo
  of 4\% is assumed for computing magnitude.}
\label{fig:cdl}
\end{figure}

To adequately cover a wide range of parameter space, two efficiency
runs were completed in which we implanted each lightcurve set with
exactly one simulated occultation event. For each event, the diameter
of the KBO was chosen randomly according to a probability, or
weighting factor, $w_D$. In the first run, objects of diameters $D
=$~0.4, 0.5, 0.6, 0.7, 0.8, and 0.9~km were added with $w_D = 1/6$ for
each diameter. In the second run, objects of diameters $D =$~1, 2, 3,
5, 10~and 30~km were added with weights $w_D = \{100, 100, 100, 30, 5,
1\} / 336$.  The modified lightcurves $k_i$ were reprocessed using the
same procedure described in \sect{sec:analysis}.  The recovered events
and the event parameters were then used to calculate the number of
expected occultation event in our survey.  That is, we calculated the
quantity
\begin{displaymath}
\aeff(D) =
  {w_D^{-1}}\sum_j\left[E_j~v_{\mathrm{rel}_j}~ H_j(D) ~/~ \Delta^2\right],
\end{displaymath}
where the sum is over all lightcurve sets where a simulated event is
successfully recovered in the reanalysis (\fig{fig:aeff}).
Essentially $\aeff(D)$ is the \emph{effective solid angle} of our
survey, insofar as TAOS can be considered equivalent to a survey that
is capable of counting every KBO of diameter $D$ in a solid angle
$\aeff$ with 100\% efficiency.

The expected number of detected events by KBOs with sizes ranging from $D_1$
to $D_2$ can then be written as
\begin{equation}
\nexp = \int\limits_{D_2}^{D_1}{dn \over dD}\aeff(D) dD,
\label{eq:nexp}
\end{equation}
where ${dn}/{dD}$ is the differential surface number density of KBOs.
The integrand of \eqn{eq:nexp} contains two factors: the
\emph{model-dependent} size distribution ${dn}/{dD}$, and the
\emph{model-independent} effective solid angle $\aeff(D)$, which
describes the sensitivity of the survey to objects of diameter $D$.
Given the observed number of events and the value of $\aeff(D)$
resulting from the efficiency calculation, we can place
model-dependent limits on the the population of KBOs. Based on the
absence of detections in this data set, any model with a size
distribution such that $\nexp~\geq~3.0$ is inconsistent with our data
at the 95\% confidence level.

Note that there are an infinite number of models that satisfy the
above requirement. We thus make the reasonable choice of a power-law
size distribution ${dn}/{dD} = n_\mathrm{B}(D/28~\mathrm{km})^{-q}$,
where $n_\mathrm{B}$ is chosen such that the cumulative size
distribution is continuous at 28~km with the results of
\citet{2004AJ....128.1364B}. We integrate \eqn{eq:nexp} from $D_2 =
28$~km down to our detection limit of $D_1 = 0.5$~km, and solve
\eqn{eq:nexp} with $\nexp = 3$, to find $q = 4.60$. Our null detection
thus eliminates any power law size distribution with $q > 4.60$ at the
95\%~c.l., setting a stringent upper limit (see \fig{fig:cdl}) to the
number density of KBOs.

\section{Conclusion}

We have surveyed the sky for occultations by small KBOs using the
three telescope TAOS system. We have demonstrated that a dedicated
occultation survey using an array of small telescopes, an innovative
statistical analysis of multi-telescope data, and a large number of
star-hours, can be used as a powerful probe of small objects in the
Kuiper Belt, and we are thus able to place the strongest upper bound
to date on the number of KBOs with $0.5~\mathrm{km} < D < 28$~km.
We continue to operate TAOS, soon with an additional
telescope, and will report more sensitive survey results in the
future.

\acknowledgements Work at NCU was supported by the grant NSC
96-2112-M-008-024-MY3. Work at the CfA was supported in part by the
NSF under grant AST-0501681 and by NASA under grant NNG04G113G. Work
at ASIAA was supported in part by the thematic research program
AS-88-TP-A02. Work at UCB was supported by the NSF under grant
DMS-0405777.  Work at Yonsei was supported by the KRCF grant to Korea
Astronomy and Space Science Institute.  Work at LLNL was performed
under the auspices of the U.S. DOE in part under Contract
W-7405-Eng-48 and Contract DE-AC52-07NA27344. Work at SLAC was
performed under U.S. DOE contract DE-AC02-76SF00515.  Work at NASA
Ames was funded by NASA/P.G.\&G.


\end{document}